\documentclass[12pt,aps,prb,preprint]{revtex4}   

\usepackage{amsmath}    
\usepackage{amsfonts}  
\usepackage{amssymb}

\begin{document}

\title{Demonstration of the spin-statistics connection in elementary quantum mechanics}

\author{J. A. MORGAN}
\affiliation{The Aerospace Corporation, P. O. Box 92957 \\Los Angeles, CA 90009, 
United States of America \\
\email{john.a.morgan@aero.org}
}


\begin{abstract}
A simple demonstration of the spin-statistics connection is presented.  The 
effect of exchange and space inversion operators on two-particle states is reviewed.  The connection 
follows directly from successive application of these operations to the two-particle wave function 
for identical particles in an $s$-state, evaluated at spatial coordinates $\pm\mbox{\bf{x}}$,
but at equal time, \emph{i.e.}, at spacelike interval.  


\end{abstract}


\maketitle

\section{Introduction}

The connection between spin and statistics, first conjectured by Pauli, and subsquently proved by
Pauli~\cite{Pauli1940}, Burgoyne~\cite{B1958}, L\"{u}ders and Zumino~\cite{LZ1958}, and others, has 
an understandable appeal to students of physics as an example of a phenomenon arising from quantum 
mechanics and relativity that has palpable consequences in the realm of everyday 
experience~\cite{DuckS1998,DuckS1997,Neuen1994}.
This paper presents a demonstration of the spin-statistics connection by an simple argument 
involving symmetry of two-particle wave functions under the combined operations of exchange and 
parity.  
It is intended to be accessible to final-year undergraduate students of quantum mechanics, 
who will have had exposure to simple angular momentum theory, the Pauli principle, and the concept 
of parity.  

It is well-known that, while relativistic quantum theory supplies sufficient 
conditions for validity of the spin-statistics connection, the question of just how
weak the necessary conditions can be remains open. That question is not addressed here.  
The demonstration given here renders in (largely) elementary language a proof 
a proof originally devised for $(j,0)$ 
or $(0,j)$ irreducible representations of the Poincar\'{e} group, sometimes called Weinberg 
fields~\cite{Morgan2005,Morgan2006,Weinberg1964,Weinberg1995}.
It exploits the properties of two-particle states constructed from identical noninteracting 
states of massive particles corresponding to Weinberg fields.  These 
single-particle 
states have the simple, definite symmetries required for the following argument:  They are 
irreducible representations of the rotation group, possess definite intrinsic parity, and satisfy 
local commutativity. 

\section{Background}

In the following, a (single-particle) state may be described by the ket vector $| \phi \rangle$
or by the wave function $\phi(\mbox{\bf{x}},t)=\langle \mbox{\bf{x}},t| \phi \rangle$.
The proof which follows concerns symmetries of two-particle wave functions  
evaluated at 
a pair of spacetime positions lying at at spacelike interval from one another.  It is possible, 
therefore, to 
specify a Lorentz frame in which the coordinates occur at equal time.
The dependence upon time will usually not be shown.

\subsection{Quantum states of higher spin}

We start by reciting results from the theory of angular momentum in quantum
mechanics that find use in the following.  
The total angular momentum operators $J_{i}$ give rise to infinitesimal rotations of a state
about the $x_{i}$ axes~\cite{FIIIJz}.
Eigenstates of total angular momentum $\hbar j$
can take on a range of values for the $z$-projection of
angular momentum~\cite{FIIIm},
\begin{equation}
\langle m| J_{z} | \phi_{j} \rangle = m \hbar \langle m| \phi_{j} \rangle,
\end{equation}
where the magnetic quantum number $m$ has values in the range~\cite{FIIm&m}
\begin{equation}
-j\leq m \leq j.
\end{equation}

Coupling of two single-particle states to a state of specified angular momentum is accomplished
with a unitary transformation whose matrix elements are Clebsch-Gordan 
coefficients~\cite{FIIIClebsch}. The Clebsch-Gordan coefficient coupling two states with total and 
magnetic angular momentum quantum numbers $(j_{a},m_{a})$ and $(j_{b},m_{b})$, respectively, to a 
state with quantum numbers $(J,M)$ is denoted 
$\langle j_{a} m_{a} j_{a} m_{a} | J M \rangle$.
Thus,
\begin{equation}
|J M \rangle=\sum_{-j_{a} \le m_{a} \le j_{a};-j_{b} \le m_{b} \le j_{b}}
\langle j_{a} m_{a} j_{b} m_{b} | J M \rangle |j_{a} m_{a};j_{b} m_{b} \rangle.
\end{equation}

\subsection{The exchange operator}

The exchange operator $\mathcal{X}$ acting on the state
\begin{equation}
| \psi \rangle = | \phi(1) \rangle | \phi(2) \rangle
\end{equation}
gives~\cite{Dirac1958,FIIIX} 
\begin{equation}
\mathcal{X} | \phi(1) \rangle | \phi(2) \rangle = | \phi(2) \rangle | \phi(1) \rangle
\end{equation}
It is assumed~\cite{footnote1} the state $| \psi \rangle$ is either symmetric (\emph{bosonic}) or 
antisymmetric 
(\emph{fermionic}) under exchange of $| \phi(1) \rangle$  and $| \phi(2) \rangle$,
\begin{equation}
\mathcal{X} | \psi \rangle = \pm | \psi \rangle.
\end{equation}
Consider now the inverse to $\mathcal{X}$.  Given $| \psi \rangle$ and another two-particle
state $| \xi \rangle$, their matrix element $\langle \xi| \psi \rangle$ should be left 
unchanged by application of $\mathcal{X}$ to both states:
\begin{equation}
\mathcal{X} \langle \xi | \mathcal{X} | \psi \rangle
=\langle \xi| \psi \rangle,
\end{equation}
which is readily seen to be the same as
\begin{equation}
 \langle \xi | \mathcal{X}^{\dagger} \mathcal{X} | \psi \rangle
=\langle \xi| \psi \rangle,
\end{equation}
or
\begin{equation}
\mathcal{X}^{-1}=\mathcal{X}^{\dagger}. \label{eq:inversesquared}
\end{equation}

\subsection{The parity operator}

The result
of the space inversion, or parity, operation on a spinless state $| \xi_{0} \rangle$ is~\cite{FIIIP} 
\begin{equation}
\langle \mathbf{x},t | \mathcal{P} | \xi_{0} \rangle) = \langle -\mathbf{x},t | \xi_{0} \rangle.
\end{equation}
Parity acting on position and momentum variables gives
\begin{eqnarray}
\mathbf{x} \Rightarrow - \mathbf{x} \nonumber \\
\mathbf{p} \Rightarrow - \mathbf{p}.
\end{eqnarray}
It follows that (orbital) angular momentum is unaltered by the parity operator:
\begin{equation}
\mathbf{x} \times \mathbf{p} \equiv \mathbf{L}  \Rightarrow  \mathbf{L},
\end{equation}
and that the $\mathcal{P}$ operation commutes with rotations.  In order that $\mathcal{P}$,
which may be regarded as a passive coordinate transformation, not alter the total angular
momentum of a wave function possessing both orbital and spin angular momentum degrees of
freedom, its effect on components of a state with definite, nonzero spin must likewise
be diagonal, allowing us to write 
\begin{equation}
\langle \mathbf{x},m | \mathcal{P} | \xi_{j} \rangle  
= \langle -\mathbf{x},m | \xi_{j} \rangle. \label{eq:PonXi}
\end{equation}

An eigenstate of parity obeys
\begin{equation}
\langle \mathbf{x},t | \mathcal{P} | \psi \rangle = \eta \langle \mathbf{x},t | \psi \rangle
\end{equation}
As two successive applications of the parity operation give the identity~\cite{FIIIeta}, 
\begin{equation}
\mathcal{P}^{2}=1,
\end{equation}
which implies 
\begin{eqnarray}
\eta^2=1; \\
\eta=\pm1 \label{eq:IntrinsicP}
\end{eqnarray}
for a state of definite parity.  States with $\eta=+1$ are symmetric under space inversion 
(\emph{even} parity), 
while states with $\eta=-1$ are antisymmetric (\emph{odd} parity).
Parity is a unitary operator, so we also have
\begin{equation}
\mathcal{P}^{\dagger}=\mathcal{P}^{-1}=\mathcal{P},
\end{equation} 
analogous to (\ref{eq:inversesquared}).

If the states $| \psi_{1} \rangle $ and  $| \psi_{2} \rangle $ respectively have parities $\eta_{1}$
and $\eta_{2}$, then the combined state  $| \psi_{1} \rangle | \psi_{2} \rangle $ has parity
\begin{equation}
\eta_{12}=\eta_{1}\eta_{2}. \label{parityproduct}
\end{equation}  

\section{The connection between spin and statistics} \label{proof}

The connection is proved with the aid of a wave function that is the amplitude 
for the particles in a two-particle state $|\psi \rangle$ to be a relative $s$-state:
\begin{equation}
\langle \mbox{\bf{r}}_{1},\mbox{\bf{r}}_{2}; 0 0 | \psi \rangle = 
\sum_{m} \langle j m j -m | 0 0 \rangle
\langle \mbox{\bf{r}}_{1}  m ; \mbox{\bf{r}}_{2}  -m | \psi \rangle.  \label{eq:state0}
\end{equation}
Given (\ref{eq:state0}), we are at liberty to evaluate it at $\mbox{\bf{r}}_{1}=\mbox{\bf{x}}$ 
and $\mbox{\bf{r}}_{2}=-\mbox{\bf{x}}$:
\begin{equation}
\langle \mbox{\bf{x}},-\mbox{\bf{x}}; 0 0 | \psi \rangle = 
\sum_{m} \langle j m j -m | 0 0 \rangle
\langle \mbox{\bf{x}}  m ; -\mbox{\bf{x}}  -m | \psi \rangle.  \label{eq:state}
\end{equation}
Consider the effect of exchange and space inversion operations on the wave functions appearing
in equation (\ref{eq:state}). 
We have
\begin{equation}
\mathcal{X} | \psi \rangle = \pm | \psi \rangle
\end{equation}
as the particles obey Bose ($+$) or Fermi ($-$) exchange symmetry,~\cite{FIIIPauli} and  
\begin{equation}
\mathcal{X} | \mbox{\bf{x}}  m ; -\mbox{\bf{x}}  -m \rangle = 
| -\mbox{\bf{x}}  -m ; \mbox{\bf{x}}  m \rangle  
\end{equation}
so that 
\begin{eqnarray} 
\langle \mbox{\bf{x}} m ; -\mbox{\bf{x}} -m | \psi \rangle= 
\langle \mbox{\bf{x}} m ; -\mbox{\bf{x}} -m |
 \mathcal{X}^{-1} \mathcal{X} | \psi \rangle \nonumber \\
= \pm \langle -\mbox{\bf{x}} -m ; \mbox{\bf{x}} m | \psi \rangle . \label{eq:shuffle}
\end{eqnarray}
Next, apply the parity operator to the wave function appearing on the RHS of
 (\ref{eq:shuffle}).
The state $| \psi \rangle$ is composed of products of identical single-particle
states $| \phi_{m} \rangle$.
According to  (\ref{parityproduct}) the parity of such a product must be even,
\begin{equation}
\mathcal{P} | \psi \rangle = | \psi \rangle
\end{equation}
with 
\begin{equation}
\mathcal{P} | -\mbox{\bf{x}} -m ; \mbox{\bf{x}} m \rangle = 
| \mbox{\bf{x}} -m ; -\mbox{\bf{x}} m \rangle
\end{equation}
leading to 
\begin{equation} 
\langle -\mbox{\bf{x}} -m ; \mbox{\bf{x}} m 
| \mathcal{P}^{-1} \mathcal{P} |\psi \rangle = 
\langle \mbox{\bf{x}} -m ; -\mbox{\bf{x}} m | \psi \rangle . \label{eq:intermediate}
\end{equation}
Inserting  (\ref{eq:intermediate}) into  (\ref{eq:shuffle}) gives
\begin{equation} 
\langle \mbox{\bf{x}} m ; -\mbox{\bf{x}} -m | \psi \rangle=
\pm \langle \mbox{\bf{x}} -m ; -\mbox{\bf{x}} m | \psi \rangle. \label{eq:combined}
\end{equation}

Upon substituting  (\ref{eq:combined}) into  (\ref{eq:state}), 
\begin{equation} 
\langle \mbox{\bf{x}},-\mbox{\bf{x}}; 0 0 | \psi \rangle = 
\pm \sum_{m} \langle j m j -m | 0 0 \rangle 
\langle \mbox{\bf{x}}, -m; -\mbox{\bf{x}}, m | \psi \rangle.  \label{eq:pmstate}
\end{equation}
We may invert the order of summation by replacing
$m$ with $-m'$ to get
\begin{equation}
\langle \mbox{\bf{x}},-\mbox{\bf{x}}; 0 0 | \psi \rangle =  
\pm \sum_{m'} \langle j -m' j m' |0 0 \rangle  
 \langle \mbox{\bf{x}} m' ; -\mbox{\bf{x}} -m' | \psi \rangle . \label{eq:reshuffle}
\end{equation}
At this point it is advantageous to rewrite  (\ref{eq:reshuffle}) in a suggestive
way.  The Clebsch-Gordan coefficient appearing in (\ref{eq:reshuffle}) is~\cite{Clebsch}
\begin{equation}
\langle j -m' j m' | 0 0 \rangle = \frac{(-1)^{(j+m')}}{\sqrt{2 j+1}}. \label{eq:Clebsch}
\end{equation}
Note that the quantity $j-m'$ is always an integer, and $2j-2m'$ an even integer.  We may write
\begin{eqnarray}
(-1)^{m'}=(-1)^{m'}(-1)^{2j-2m'} \nonumber \\
=(-1)^{2j}(-1)^{-m'}
\end{eqnarray}
and conclude
\begin{equation}
\langle j -m' j m' | 0 0 \rangle = (-1)^{2 j}
\langle j m' j -m' | 0 0 \rangle .
\end{equation}
Employing this relation in (\ref{eq:reshuffle}) and recalling  (\ref{eq:state}) gives us
\begin{equation}
\langle\mbox{\bf{x}},-\mbox{\bf{x}}; 0 0 | \psi \rangle 
= \pm (-1)^{2 j} \langle\mbox{\bf{x}},-\mbox{\bf{x}}; 0 0 | \psi \rangle. \label{eq:almostthere}
\end{equation}

The singlet wave function appearing in (\ref{eq:almostthere}) is nonvanishing if the 
individual wave functions 
from which it is constructed are themselves
nonvanishing.  A proof of this assertion appears in the Appendix. 
If we can assume the matrix element on both sides of 
(\ref{eq:almostthere}) does not vanish, we immediately have 
\begin{equation}
1=\pm(-1)^{2j}. \label{eq:drumroll}
\end{equation} 
According to (\ref{eq:drumroll}),
states $|\mbox{\bf{x}},-\mbox{\bf{x}} \rangle$ with $2j$ \emph{even} necessarily have 
\emph{Bose} exchange symmetry, while those with $2j$ \emph{odd}\ necessarily have 
\emph{Fermi} symmetry.  This is the connection between spin and statistics.

\section{Discussion}

The demonstration just presented is neither so simple nor rigorous as the formal proof in relativstic 
field theory given by 
Burgoyne~\cite{B1958}.  On the other hand it does rely, for the most part, upon concepts and methods 
taken from elementary quantum mechanics.
Apart from the material appearing in the appendix, it depends on
nothing that cannot readily be obtained (or at a minimum, motivated) starting from pertinent 
discussions in the Feynman Lectures.
It may appear that the proof as given in Section \ref{proof} could be accomplished without
any reliance upon relativistic quantum mechanics.
However, at certain points the argument rests
upon assumptions that flow in a natural and unforced way from requirements of relativistic
symmetry, but which would arguably enter a truly nonrelativistic exposition in neither fashion.

An instance is the symmetry of the wave function in 
(\ref{eq:shuffle}), which is a disguised statement of an 
equal-time commutation relation. 
Exhibiting the dependence upon $t$,  (\ref{eq:shuffle}) becomes
\begin{equation}
\langle (\mbox{\bf{x}},t) m ; (-\mbox{\bf{x}},t) -m | \psi \rangle 
= \pm \langle (-\mbox{\bf{x}},t) -m ; (\mbox{\bf{x}},t) m | \psi \rangle . \label{eq:r2state}
\end{equation}
In (\ref{eq:r2state}), the wave functions that give the probability amplitude 
for the particles (1) and (2) at spatial position $\pm \mbox{\bf{x}}$ are 
evaluated at equal time $t$.  Put another way, 
\begin{equation}
|\mbox{\bf{x}}_{2}-\mbox{\bf{x}}_{1} |^{2} -(t_{2}-t_{1})^2 > 0 \label{eq:lightcone}
\end{equation}
The statement that a relation holds between two points separated by nonzero distance at equal
time has no unambiguous meaning in special relativity~\cite{FIsimul}.  
Equation (\ref{eq:lightcone}), 
however, is an invariant statement under arbitrary Lorentz transformations.  In proofs of the 
spin-statistics relation, the exchange symmetry that appears in (\ref{eq:shuffle}) is normally 
stipulated subject to (\ref{eq:lightcone}).  One says that wave functions of identical particles 
commute or anticommute outside the light cone~\cite{FIlightcone}.

Moreover, it was assumed that massive particle states exist with certain simple, conjoined symmetries 
with respect to the operations of parity and rotation.  
As noted earlier, the assumed symmetries of the states are 
those of an irreducible representation of the Poincar\'{e} group.~\cite{Weinberg1964}
Thus, elements of the present demonstration that would enter a genuinely nonrelativistic proof as 
distinct hypotheses all follow from the single requirement of Poincar\'{e}
invariance in an explicitly relativistic treatment.  Granted this observation, the nonrelativistic 
view does not appear to be the parsimonious one, even should it be possible to construct a 
completely nonrelativistic proof.

\section{Appendix}

\textit{We apologize for the fact that we cannot give you an elementary explanation.}

-R. P. Feynman, Ref.~\cite{FIIIJz}, Vol. III, p. 4-3

In the following it will be convenient to write two-particle wave functions in factored form so that,
$\emph{e. g.,}$ the wave function in (\ref{eq:state}) is written as
\begin{equation}
\langle \mbox{\bf{x}} m ; -\mbox{\bf{x}} -m | \psi \rangle =
\langle \mbox{\bf{x}} m | \phi_{j}(1) \rangle \langle -\mbox{\bf{x}} -m | \phi_{j}(2) \rangle.
\end{equation}
From single-particle wave functions for spin $j$, which may be assumed to belong to an irreducible 
representation of the rotation group, form
\begin{equation}
(\xi_{j},\phi_{j}) \equiv 
(-1)^{-j} \sum_{m} \int {d^{3}x} \, \langle j m j -m | 0 0 \rangle
\langle \mbox{\bf{x}} m | \xi_{j} \rangle  
\langle \mbox{\bf{x}} m | \phi_{j} \rangle^{*}. \label{eq:IP}
\end{equation}
This quantity serves as an inner product in the Hilbert space of wave functions on 
$\mathbf{R}_{3}$~\cite{Morgan2006}.  In
\begin{equation}
(\phi_{j},\phi_{j})=
(-1)^{-j} \sum_{m} \int {d^{3}x} \, \langle j m j -m | 0 0 \rangle
\langle \mbox{\bf{x}} m | \phi_{j} \rangle  
\langle \mbox{\bf{x}} m | \phi_{j} \rangle^{*}.  \label{eq:norm}
\end{equation}
we may write
\begin{equation}
\langle \mbox{\bf{x}} m | \phi_{j} \rangle=f_{j}(r)\mathcal{Y}_{jm}(\Omega)
\end{equation}
at radius $r$. 
Here the function $\mathcal{Y}_{jm}(\Omega)$ is a suitable angular momemtum eigenfuncton that 
generalizes the properties of spherical harmonics to include half-integral as well as 
integral angular momenta~\cite{Edmonds1960,FanoRacah1959}.
It may be defined so as to share with ordinary
spherical harmonics $Y_{lm}(\Omega)$ the conjugation property
\begin{equation}
\mathcal{Y}^{*}_{jm}=(-1)^{m}\mathcal{Y}_{j \, -m}.
\end{equation}
 We also have
\begin{eqnarray}
\int {d\Omega} \mathcal{Y}^{*}_{jm'}\mathcal{Y}_{jm}= && 
\int {d\Omega} \mathcal{Y}_{jm'}\mathcal{Y}^{*}_{jm}=\delta_{m'm}.
\end{eqnarray}

The  angular momentum ladder operators $J_{\pm}$ are defined by
\begin{equation}
J_{\pm}=J_{x} \pm iJ_{y}
\end{equation}
and have the effect of raising and lowering $m$:
\begin{equation}
\langle m | J_{\pm} | \phi_{j} \rangle =
-i \hbar \sqrt{(j\mp m)(j\pm m+1)} \langle m \pm 1 | \phi_{j} \rangle \label{eq:ladderopdef}
\end{equation}
The $J_{\pm}$ are differential operators that
act on orbital and spin degrees of freedom $\emph{only}$~\cite{Edmonds1960}.
This observation means that the $J_{\pm}$ raise and lower $m$ in $\mathcal{Y}_{jm}(\Omega)$
and have no effect upon $f_{j}(r)$. The radial weight $f_{j}(r)$ can, therefore, have no dependence 
upon $m$~\cite{Baym69}.
Recalling the definition of the Clebsch appearing in (\ref{eq:norm}) 
(\emph{vide.}  (\ref{eq:Clebsch})), we find
\begin{equation}
(\phi_{j},\phi_{j}) = \int {r^{2}dr} \, f_{j}(r) f_{j}^{*}(r) \ge 0,
\end{equation}
with equality iff $f_{j}(r)$ vanishes everywhere.
Should
\begin{equation}
\sum_{m} \langle j m j -m | 0 0 \rangle
\langle \mbox{\bf{x}} m | \phi_{j} \rangle  
\langle \mbox{\bf{x}} m | \phi_{j} \rangle^{*}=0, \forall \, \mbox{\bf{x}}
\end{equation} 
then $(\phi_{j},\phi_{j})$ will vanish. But $(\phi_{j},\phi_{j})=0$  
iff $\langle \mbox{\bf{x}} m | \phi_{j} \rangle$ vanishes, 
as well. 

Assume $| \zeta_{j} \rangle$ is a state of a spin $j$ particle such that
\begin{equation}
\langle \mbox{\bf{x}} m | \zeta_{j} \rangle \ne 0
\end{equation} 
From $| \zeta_{j} \rangle$ form
\begin{equation}
\langle \mbox{\bf{x}} m |\phi_{j} \rangle \equiv
\langle \mbox{\bf{x}} m | \zeta_{j} \rangle^{*}
\pm \langle -\mbox{\bf{x}} -m | \zeta_{j} \rangle.  \label{eq:xidef}
\end{equation}
Then
\begin{eqnarray}
\sum_{m} \langle j m j -m | 0 0 \rangle \langle \mbox{\bf{x}} m |\phi_{j} \rangle
\langle -\mbox{\bf{x}} -m |\phi_{j} \rangle= \nonumber \\ 
\pm \sum_{m} \langle j m j -m | 0 0 \rangle
\langle \mbox{\bf{x}} m |\phi_{j} \rangle
\langle \mbox{\bf{x}} m |\phi_{j} \rangle^{*}. \label{eq:xisprods}
\end{eqnarray}
As a general rule, the wave function $\langle \mbox{\bf{x}} m |\phi_{j} \rangle$ will have 
nonvanishing norm and the RHS of (\ref{eq:xisprods})
will differ from zero.  But suppose that for one choice of sign in (\ref{eq:xidef}), 
$\langle \mbox{\bf{x}} m |\phi_{j} \rangle$ 
were to vanish  $\forall \, \mbox{\bf{x}}$.
In that event $\langle \mbox{\bf{x}} m |\phi_{j} \rangle$, and hence (\ref{eq:xisprods}), cannot 
vanish for the other choice.  
We suppose in the main text that the appropriate choice of sign has been made, if necessary, and 
that (\ref{eq:state}) is therefore nonvanishing on some open set of $\mbox{\bf{x}}$.

\end{document}